Oral # 0321

# Real-Time Phase Contrast MRI to quantify Cerebral arterial flow change during variations breathing

Pan Liu[1], Sidy Fall[1], Serge Metanbou[2], Olivier Baledent[1,3]

[1]CHIMERE UR 7516, Jules Verne University, Amiens, France.

[2]Department of Radiology, Jules Verne University Hospital, Amiens, France

[3]Medical Image Processing Department, Jules Verne University Hospital, Amiens, France.

## Summary of Main Findings (250 characters, 249/250)

Cerebral arterial blood flow (CABF) can be investigated in few seconds without any synchronization by Real-Time phase contrast. Significant changes of CABF were found between expiration and inspiration during normal breathing of healthy volunteers.

## Synopsis (100/100)

Real-time phase contrast MRI has been applied to investigate cerebral arterial blood flow (CABF) during normal breathing of healthy volunteers. We developed a novel time-domain analysis method to quantify the effect of normal breathing on several parameters of CABF. We found the existence of a delay between the recorded respiratory signal from the belt sensor and the breathing frequency component presents in the reconstructed arterial blood flows. During the expiratory, the mean flow rate of CABF increased by 4.4±1.7%, stroke volume of CABF increased by 9.8±3.1% and the duration of the cardiac period of CABF increased by 8.1±3%.

## Introduction (850 words to conclusion 849/850)

The effect of respiration on cerebral arterial blood flow is not very well understood[1,2]. Conventionally, cardiac gated phase contrast sequences are used for the quantification of cerebral blood flow[3-6], however, it cannot be applied to studies in this area because it only provides an average Cardiac Cycle Flow Curve (CCFC)[7].

With real-time phase contrast (RT-PC) sequences, it has currently been shown that there is an effect of deep breathing or coughing on cerebral blood flow and cerebrospinal fluid[8-11]. However, it is currently not well understood whether and to what extent normal respiration affects cerebral arteries blood flow, in this study, we used RT-PC to quantify this effect using a novel time-domain analysis method.

## Methods

– **Image acquisition**

10 participants (19 ~ 35 age; 4 female) were examined using a clinical 3T scanner, maximum gradients 80 mT/m Rate of gradient increase 120 mT m$^{-1}$ ms$^{-1}$ and a 32 channels head coil.

The RT-PC used in this study was a multi-shot, gradient-recalled echo-planar imaging sequence with a Cartesian trajectory. Using the parameter presented in Figure 1, the time resolution was 75ms/image, a respiratory monitoring belt was used for recording the respiratory signal. Participants were not asked to perform a specific respiratory pattern during the data collection. RT-PC was performed for each participant to quantify CABF at the intracranial and extracranial levels (Figure 1).

– **Image Processing**

The images of RT-PC were processed with a homemade software – Flow[12] to extract the CABF signal. The region of interest (ROI) was defined by a semi-automatic dynamic segmentation algorithm and can be also corrected manually. The software then used an automatic background field correction algorithm to find the stationary tissue around the ROI, thus correcting for velocity errors due to eddy currents. Anti-aliasing algorithm are also supported in the software. After the segmentation and calibration, the continuous flow rate signal of RT-PC was obtained (Figure 1. E).

$CABF_{extra}$ was obtained by summing arterial flows in the extracranial level, and $CABF_{intra}$ was obtained by summing arterial blood flows in the intracranial level.

- **Quantification of the effect of respiration on arterial blood flow**

The quantification process is shown in Figure 2. **Signal segmentation:** the software automatically segmented the continuous flow rate signal into multiple independent CCFCs. In each CCFC three parameters were calculated (mean flow, stroke volume, cardiac period). **Definition of respiratory interval:** the CCFCs were labelled as in expiratory interval or in inspiratory interval based on the respiratory signal. Two average expiratory and inspiratory CCFCs were calculated and used to compare the two respiratory periods. ***$Diff_{Ex-In}$ (parameter, delay) signal***: the percentage difference $Diff_{Ex-In}$ was calculated for each parameter between expiratory CCFC and inspiratory CCFC. We hypothesis that respiration impacts blood with a possible delay between the flow curves and the signal recorded by the respiratory belt sensor. This delay was calculated by applying a phase shift on the respiratory interval that produce the maximum $Diff_{Ex-In}$ for the considered parameter of interest. **Extraction results,** use the maximum value of $Diff_{Ex-In}$ *(parameter, delay)* to indicate the intensity and *delay* of the effect of respiration on this *parameter* of CABF. Also, the ratio of the corresponding *delay* to the average respiratory cycle time was calculated and recorded as *delay%*.

- **Statistical analysis**

Some arterial flows were excluded due to the poor quality of the signal acquisition. Finally, quantification of the effect of respiration was performed on 58 arteries. The correlation between $Diff_{Ex-In}$ and *delay%* of each parameter was also tested using Spearman's correlation test.

## Results

Figure 3 shows the $Diff_{Ex-In}$ *(parameter, delay)* signal of the carotid arteries of two participants.

Figure 4 shows the results of CABF quantification and respiratory impact quantification. The cerebral blood flows were equal to 740±74 ml/min at the extracranial level and 679±119 ml/min at the intracranial level. The $Diff_{Ex-In}$ were significantly smaller without considering the delay. In the case of considering delay, the *delay%* and $Diff_{Ex-In}$ of mean flow rate are smaller than that of both stroke volume and cardiac period ($p < 0.01$). There was also a correlation between the *delay%* and $Diff_{Ex-In}$ of mean flow rate and of cardiac period (Figure 5).

## Discussion

In this study, the cerebral arterial blood flow obtained using RT-PC was accordance with the results previously obtained using cardiac gated phase contrast[4,8]. It was also found that under normal respiration, cerebral artery blood flow increased by 4.4±1.7% during the expiratory interval compared to the inspiratory interval.

With increasing *delay%*, the $Diff_{Ex-In}$ of the cardiac period was positively correlated, while the $Diff_{Ex-In}$ of the mean flow was negatively correlated, which is a very interesting phenomenon and seems to be a positive response to stabilize stroke volume. This can be further confirmed in a subsequent study.

## Conclusion

Although there are some artifactual effects, RT-PC is very useful for investigating cerebral blood flow in real time within a few seconds, without any synchronizer.

RT-PC combined with this novel method allows us to quantify the effect of respiration on multiple parameters of blood flow or cerebrospinal fluid, and the *delay%* parameter provides a new analytical perspective. It provides a technical support for subsequent studies on the effect of respiration on cerebral fluids circulation.

## Figures (5/5)

Oral # 0321

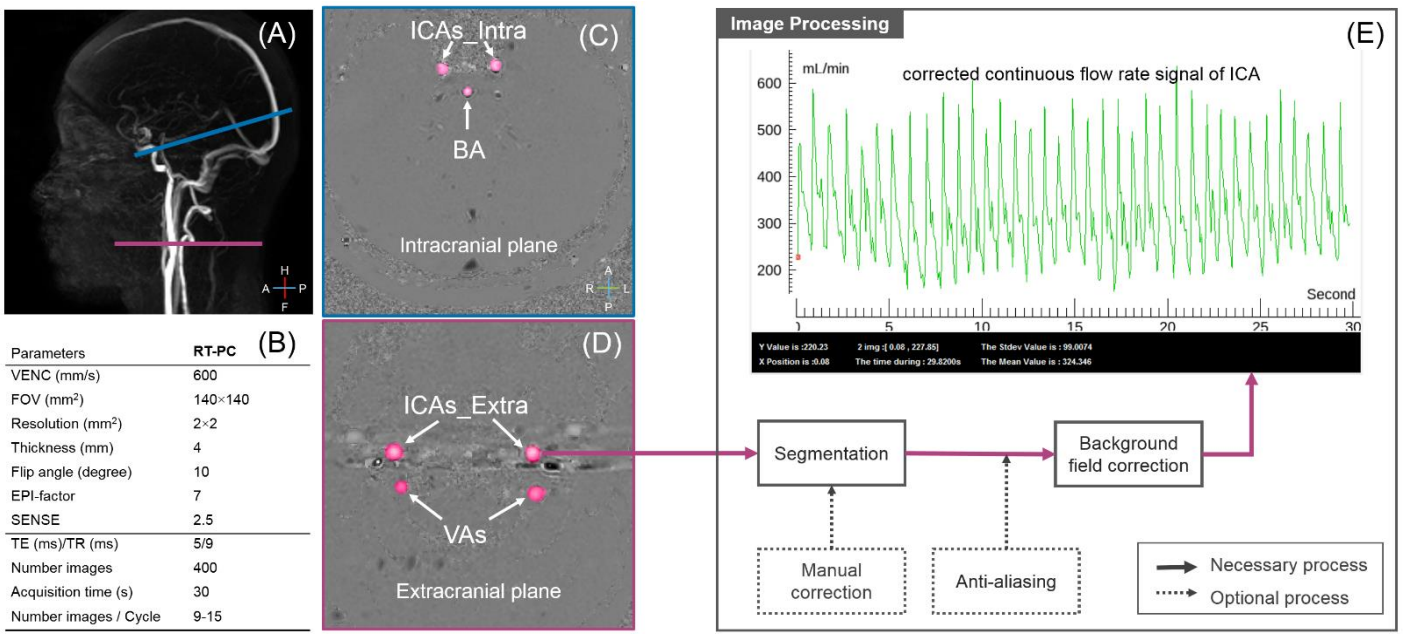

Figure 1: **Location of the acquisition planes (A) and arteries (C&D), protocol of RT-PC (B) and the image processing flowchart (E).** Right & left internal carotid artery (ICA) and right & left vertebral artery (VA) in the extracranial acquisition plane; right & left internal carotid and basilar artery (BA) in the intracranial acquisition plane.

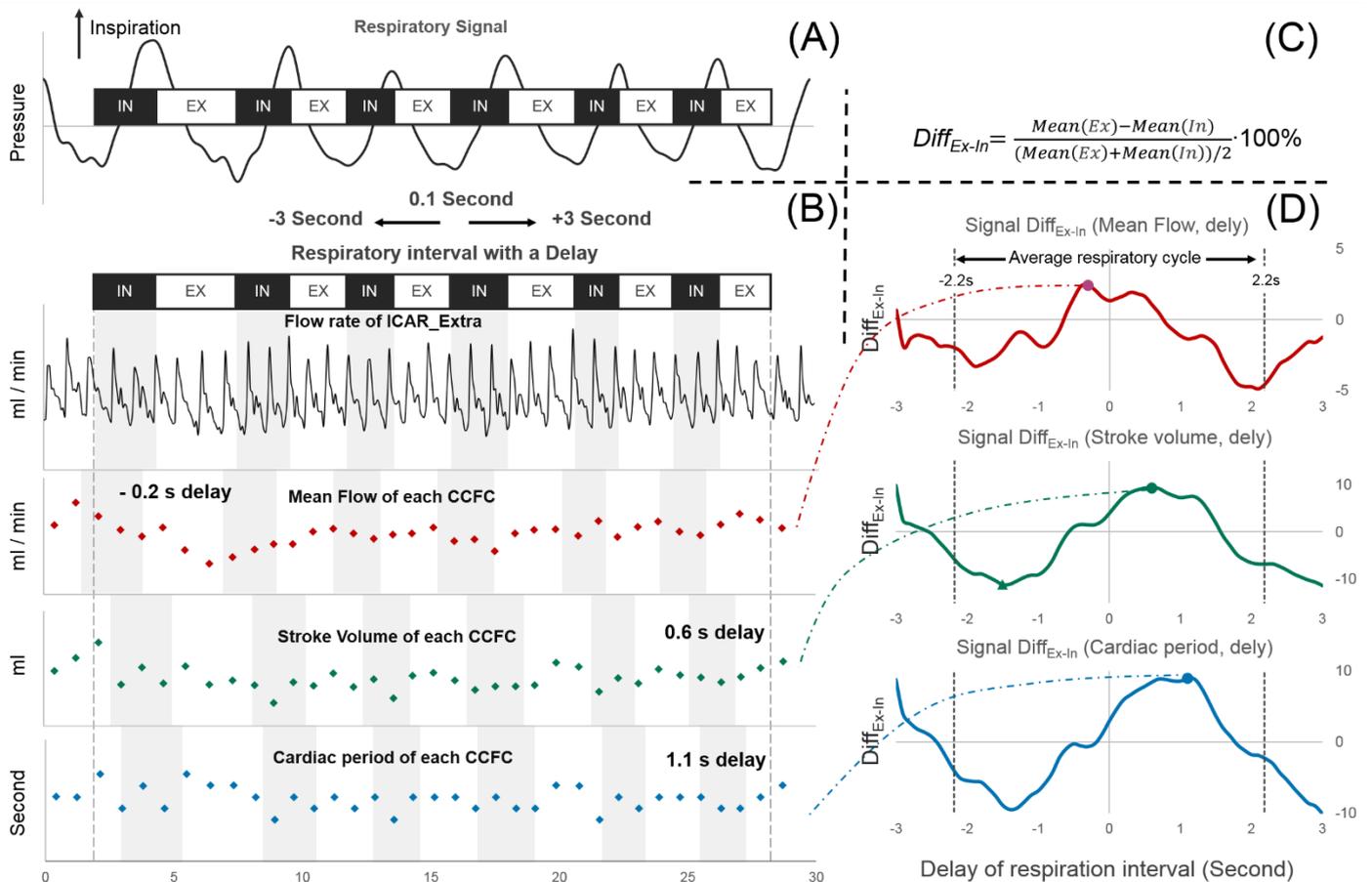

Figure 2: **Schematic diagram of the quantitative respiratory impact process.** Define inspiratory interval (IN) and expiratory interval (EX) using respiratory signal (A). The flow rate signal is segmented into multiple independent CCFCs and the three parameters are extracted (B), then the percentage difference between expiratory CCFC and inspiratory CCFC ($Diff_{Ex-In}$) is calculated for each parameter by equation C. Finally, the maximum $Diff_{Ex-In}$ can be found by shifting the respiratory interval (*delay*) (D).

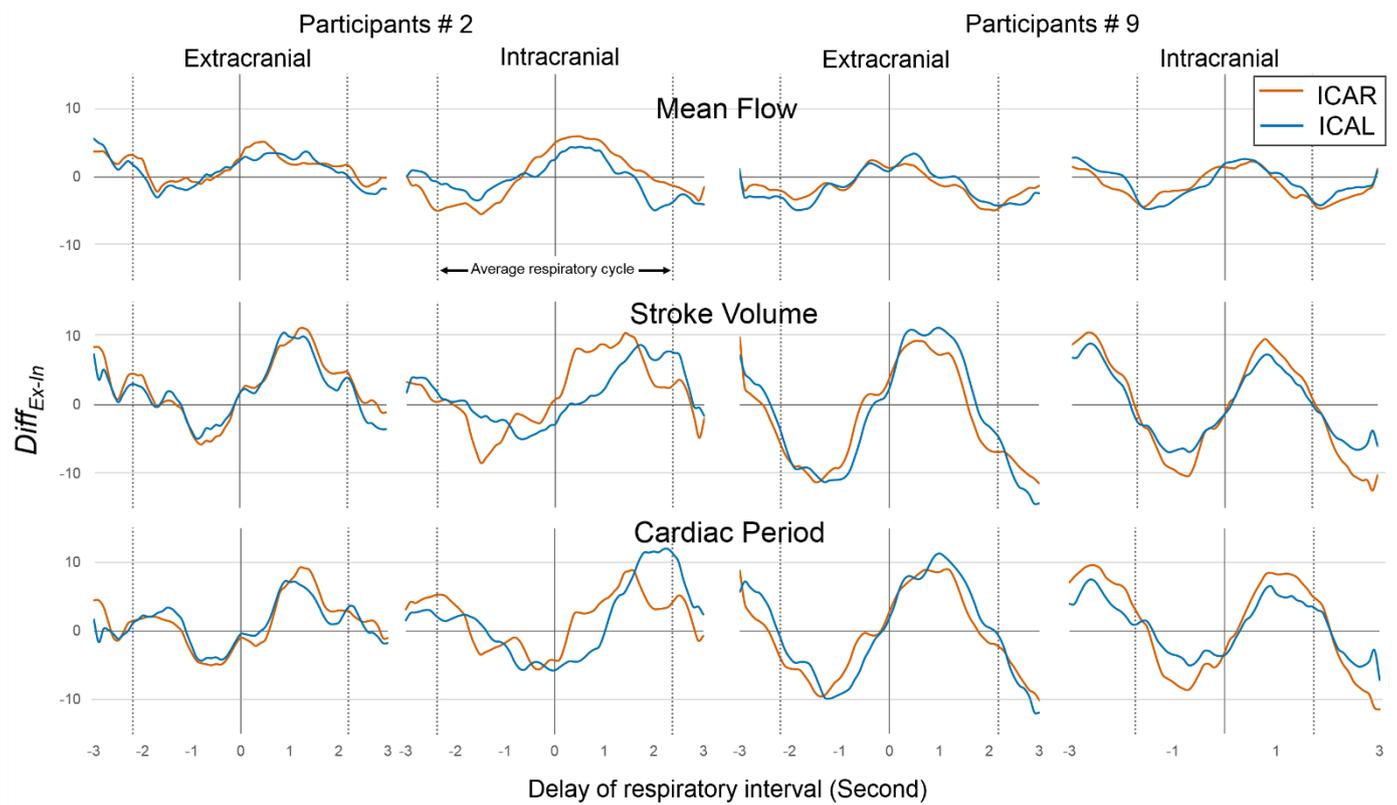

Figure 3: **Example of $Diff_{Ex-In}$ (parameter, delay) signal of the internal carotid arteries in two acquisition planes from two participants.** ICAR indicates right internal carotid artery, the *parameter* indicates mean flow rate, stroke volume or cardiac period. The dotted line indicates an average respiratory cycle.

|  | Cerebral arterial blood flow in | | $Diff_{Ex-In}$ without delay | $Diff_{Ex-In}$ with delay | |
|---|---|---|---|---|---|
|  | Extracranial (ICAs+VDs) | Intracranial (ICAs+BA) | $Diff_{Ex-In}$ | $Diff_{Ex-In}$ | Delay% |
| Mean flow (ml/min) | 740±74 | 679±119 | 2.7±2.3 (%) | 4.3±1.9 (%) | 13±12** |
| Stroke volume (ml) | 11.7±1.8 | 11.0±2.4 | 0.5±3.9 (%) | 10±3.7 (%) | 27±9.5 |
| Cardiac period (second) | 0.94±0.13 | 0.96±0.14 | -2.1±3.4 (%) | 8.4±3.4 (%) | 28±9.6** |

Five participants containing quantitative data for all seven arteries were selected for comparing the blood flow parameters in the two acquisition planes. The duration of respiratory cycle is 4.3±1 second, ICA indicates internal carotid artery, VA indicates vertebral artery, BA indicates basilar artery. $Diff_{Ex-In}$ indicates the percentage difference between expiratory CCFC and inspiratory CCFC. *Delay%* Indicates the ratio of delay to respiratory cycle duration.

Figure 4: **Quantification of cerebral arterial blood flow and respiratory effects in 5 participants.** The results of the quantification of respiratory effects include both cases, without and with consideration of delays. Using Spearman's test to detect the correlation between *delay%* and $Diff_{Ex-In}$. ** indicted $p < 0.01$.

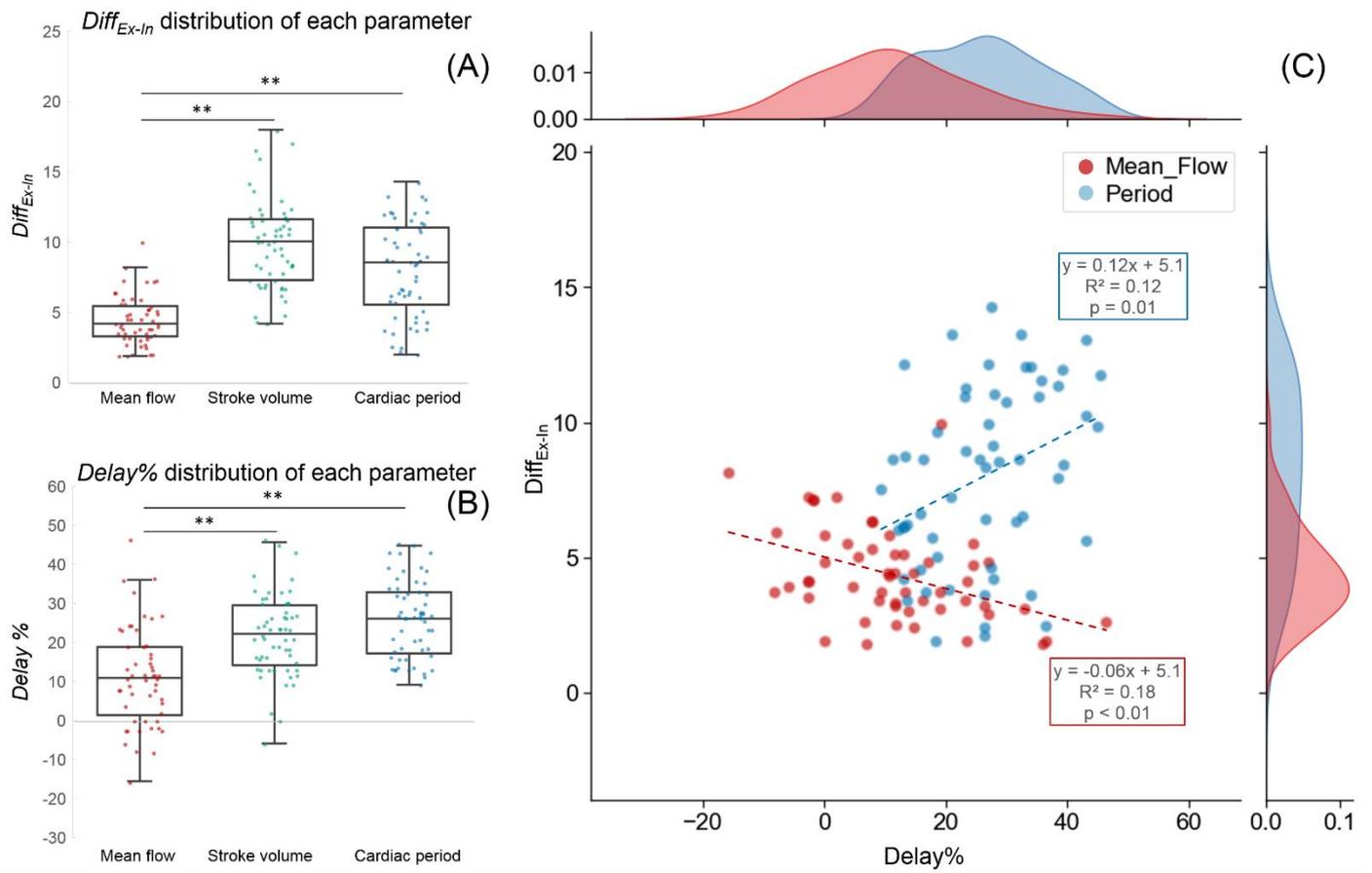

Figure 5: **Distribution plots of $Diff_{Ex-In}$ (A) and *delay%* (B) for each parameter.** There is a negative correlation between the $Diff_{Ex-In}$ and *Delay%* of the mean flow rate and a positive correlation between the $Diff_{Ex-In}$ and *Delay%* of the cardiac period (C). ** indicates significant difference ($p < 0.01$, Wilcoxon signed-rank test).

Oral # 0321

# Acknowledgements

This research was supported by EquipEX FIGURES (Facing Faces Institute Guilding Research), European Union Interreg REVERT Project, Hanuman ANR-18-CE45-0014 and Region Haut de France.

Thanks to the staff members at the Facing Faces Institute (Amiens, France) for technical assistance.

Thanks to David Chechin from Phillips industry for his scientific support.